\documentclass[a4paper,fleqn]{cas-sc}

\usepackage[numbers,sort&compress]{natbib}
\usepackage[T1]{fontenc}
\usepackage{booktabs}
\usepackage{tabularx}
\PassOptionsToPackage{hyphens}{url}
\usepackage{url}

\begin{document}
\let\WriteBookmarks\relax
\def\floatpagepagefraction{1}
\def\textpagefraction{.001}

\shorttitle{Accent Conversion Survey}
\shortauthors{Halychanskyi et al.}

\title[mode=title]{Accent Conversion: A Problem-Driven Survey of Sociolinguistic and Technical Constraints}

\author[1,2]{Yurii~Halychanskyi}
\cormark[1]
\ead{yuriih2@illinois.edu}
\credit{Conceptualization, Methodology, Investigation, Writing -- original draft, Writing -- review \& editing}

\author[3]{Jianfeng~Steven~Guo}
\ead{jg31@illinois.edu}
\credit{Writing -- original draft, Writing -- review \& editing}

\author[1,2]{Volodymyr~Kindratenko}
\ead{kindrtnk@illinois.edu}
\credit{Supervision, Writing -- review \& editing}

\cortext[1]{Corresponding author}

\affiliation[1]{organization={Siebel School of Computing and Data Science, University of Illinois Urbana-Champaign},
            city={Urbana},
            state={IL},
            country={USA}}

\affiliation[2]{organization={National Center for Supercomputing Applications, University of Illinois Urbana-Champaign},
            city={Urbana},
            state={IL},
            country={USA}}

\affiliation[3]{organization={Department of East Asian Languages and Cultures, University of Illinois Urbana-Champaign},
            city={Urbana},
            state={IL},
            country={USA}}

\begin{abstract}
Accent conversion has rapidly progressed alongside growing interest in improving global cross-cultural communication. This survey presents an overview of the evolution of accent conversion methodologies, analyzing how the field has developed in response to fundamental challenges related to data alignment, representation disentanglement, and resource scarcity. We trace the progression from early rule-based digital signal processing approaches such as spectral manipulation and formant-based analysis to modern neural architectures capable of flexible and reference-free accent transformation. In addition, the survey situates accent conversion within its linguistic foundations and examines how different application requirements impose varying constraints on the balance between accent modification and speaker identity preservation. Finally, it reviews commonly used speech datasets and evaluation methodologies, identifies persistent challenges, and outlines directions for future research aimed at achieving more controllable and perceptually consistent accent conversion.
\end{abstract}


\date{}

\maketitle

\section{Introduction}
\label{sec:intro}
Accent is defined by the segmental and suprasegmental features of a speaker's phonetics and phonology, that is, how a person pronounces words when speaking \cite{Trudgill2006}. Together with voice quality, accent constitutes a fundamental component of speaker identity. In particular, accent is often closely linked to a speaker's group identity, especially with respect to geographic origin and social class \cite{Trudgill2006, markl2023everyone}.
 Accent conversion (AC) refers to modifying a speaker's accent toward a target accent while preserving other perceptual identity attributes \cite{17}.

 In the context of global communication, where languages such as English often serve as international lingua francas among speakers from different linguistic backgrounds, accent differences can result in intelligibility challenges as well as social biases and stereotypes.
 For instance, speakers with foreign accents are often subject to negative bias from adult listeners, who tend to perceive them as less trustworthy, less educated, less intelligent, and less competent compared to native-accented speakers \cite{lorenzoni2024does}. Furthermore, contemporary speech technologies, including automatic speech recognition (ASR) and text-to-speech (TTS) systems, are typically biased toward high-resource accents (e.g., Standard American English), leading to suboptimal performance for speakers with underrepresented or low-resource accents \cite{21}. Consequently, AC has gained substantial attention as a tool to enhance intelligibility, mitigate accent-related bias, and improve equitable access to speech technologies for speakers of underrepresented accents.

Despite significant advances in accent conversion techniques, surveys that frame the field through its core sociolinguistic foundations and scientific constraints remain scarce. While prior surveys (e.g., \cite{competitor2025, competitor2024}) provide an engineering-centric overview of model architectures and feature extraction pipelines, this work integrates two complementary perspectives. First, it grounds accent conversion in linguistic foundations from phonetics and phonology as well as sociolinguistics, treating accent as both a structured pattern of segmental and suprasegmental realization and a salient social signal that can trigger stereotypes, bias, and unequal technology performance across speaker groups. Second, it adopts a constraint-driven view of methodological progress, analyzing how accent conversion paradigms emerged as responses to persistent limitations in supervision, representation, and data availability. In addition, it examines how different application requirements impose varying constraints on the balance between accent modification and speaker identity preservation.

Accordingly, this survey organizes the literature around a constraint-driven framework that examines how accent conversion methods have evolved in response to three recurring scientific bottlenecks:
(i) the challenge of learning accent mappings without reliable utterance-level parallel data, which requires establishing correspondence between source and target speech through explicit or implicit alignment,
(ii) the reference-free requirement that motivates disentangling accent from linguistic content and speaker identity,
and (iii) persistent data scarcity and imbalance for low-resource accents that limits scalability and generalization.
By organizing prior work around these constraints rather than model families alone, this survey highlights not only how accent conversion systems are built, but also why specific methodological shifts have emerged and what new trade-offs each shift introduced for accent similarity, intelligibility, naturalness, and identity preservation.

The remainder of the paper is organized as follows. Section~\ref{sec:accent-linguistics} defines accent and its linguistic foundations and elaborates on the importance of accent studies.
Section~\ref{sec:ac-def} introduces AC, discusses its relation to speaker identity, and illustrates practical applications.
Section~\ref{sec:ac-challenges} identifies primary challenges in AC.
Section~\ref{sec:related-tasks} relates AC to other speech processing tasks.
Section~\ref{sec:evaluation} reviews common evaluation methodologies.
Section~\ref{sec:datasets} presents an overview of relevant speech datasets.
Section~\ref{sec:taxonomy} provides a historical, constraint-driven taxonomy of AC methods.
Section~\ref{sec:future} explores promising future research directions.
Finally, Section~\ref{sec:conclusion} concludes the survey.

\section{Accent in Linguistics Research}
\label{sec:accent-linguistics}

The study of accents is a fundamental aspect of linguistics, which investigates the structure, acquisition, and use of language \cite{MESTHRIE2006472}. Linguistics includes several subfields that contribute significantly to understanding accents. \textit{Phonetics} and \textit{phonology}, which analyze the physical properties of speech sounds and the systematic organization of sounds in languages, help to identify features that characterize different accents. \textit{Sociolinguistics}, which studies the influence of social factors on language, examines how accents reflect identity and group membership. \textit{Applied linguistics}, which is concerned with the practical uses of linguistic knowledge, such as in language education and speech therapy, seeks to understand how to support effective communication.

\subsection{Defining Accent}

Accent refers to group-level patterns of pronunciation at both segmental and suprasegmental levels, encompassing how words are realized phonetically and prosodically within a given speech community \cite{Trudgill2006}.

Every speaker has an accent \cite{wardhaugh2021introduction}. Accents are commonly classified as either standard or non-standard \cite{Giles2004}. The standard accent is typically associated with higher socioeconomic status, institutional authority, and widespread media usage within a given community. For instance, in British English, Received Pronunciation (RP) is frequently regarded as the standard variety \cite{Giles2004}. In contrast, non-standard accents are often perceived as foreign or associated with minority groups or lower socioeconomic status \cite{fuertes2012meta}. A foreign accent is often taken as an indicator of non-nativeness because second-language learners tend to carry over phonetic features from their first language \cite{Trudgill2006}.

Accents manifest through both segmental (consonants and vowels) and suprasegmental dimensions, including pitch, rhythm, and speaking rate \cite{Trudgill2006}. At the segmental level, for example, voice onset time (VOT), defined as the interval between the release of a stop consonant and the onset of the following vowel, is one cue that contributes to the perception of a foreign accent \cite{schoonmaker2015measuring}. At the suprasegmental level, features such as reduced tonal space, slower speech rate, and longer pauses have also been shown to influence perceived accent, as observed, for example, in heritage Cantonese \cite{kan2020suprasegmental}. Together, these segmental and suprasegmental cues highlight the multi-dimensional nature of accent, underscoring the need for accent conversion systems to model both fine-grained phonetic realizations and broader prosodic patterns.

Accent is not the same as dialect. While accent concerns pronunciation patterns, a dialect encompasses a broader range of linguistic features, including pronunciation, vocabulary, and grammar \cite{Trudgill2006}. Grammar governs the rules and structure of sentences, vocabulary refers to the set of words used in a language or dialect, and spelling pertains to the written form of words. Although these aspects are crucial for language modeling and text processing, accent conversion methods covered in this survey primarily target acoustic features associated with pronunciation patterns.

\subsection{The Importance of Studying Accents}

Understanding accents has multiple important implications. \textbf{Language variation}: accents reveal patterns of language change, contact, and migration, linking phonetic realization to social and geographic history. \textbf{Communication and learning}: insights from accent research inform pronunciation teaching in second-language acquisition, improving intelligibility and learner outcomes. \textbf{Social identity and perception}: accents shape social evaluation in education, employment, and everyday interaction, with documented effects on attitudes and bias. \textbf{Cognitive and psychological aspects}: findings on accent perception and processing contribute to theories of speech perception, category learning, and adaptation.

\section{Accent Conversion}
\label{sec:ac-def}

\subsection{Defining Accent Conversion}

Accent conversion refers to the process of modifying a speaker's accent to resemble a target accent while preserving the perceptual identity of the original speaker \cite{17}. In addition to speaker identity, certain approaches may optionally preserve other source utterance attributes, such as emotion or speaking intention, depending on the target application. A specialized subtask of AC is Foreign Accent Conversion (FAC), which specifically targets the transformation of non-native accents into native-sounding ones.

Speaker identity comprises the set of acoustic characteristics that make a speaker's voice recognizable and distinct, encompassing both individual traits such as voice quality and group-level traits such as accent \cite{new_laver_voice_quality,new_kreiman_sidtis_voice,new_lavan2019_flexible}. Although accent itself constitutes part of speaker identity, altering it does not necessarily eliminate the perception of the same speaker. In practice, listeners can still recognize a speaker even when some identity cues such as accent are modified, provided that other stable and anatomically grounded traits, like voice quality, remain preserved. Therefore, AC seeks to maintain these perceptually salient cues that contribute most strongly to speaker recognizability. Voice quality can be divided into phonation and timbre. Phonation describes how the vocal folds vibrate at the glottal source, giving rise to traits such as breathy, creaky, or pressed voice qualities, whereas timbre arises from vocal-tract filtering that shapes the perceived color of the voice (for example, nasal, bright, or dark) \cite{new_laver_voice_quality,new_kreiman_sidtis_voice}.

\subsection{AC Applications}

The degree to which specific accent features, such as phoneme articulation, intonation, or overall speaking rate, are modified depends heavily on the intended application. It involves a deliberate trade-off between accent transformation and speaker identity preservation. For instance, in language learning applications, AC is often used to create "golden speaker" models that prioritize fully native-like segmental and suprasegmental features. Even if this reduces traces of the original speaker's identity, the goal is to provide learners with a clear, native-like utterance target \cite{2}. However, learners may also benefit from selective adaptation, focusing on specific accent elements such as intonation patterns or phoneme articulation without completely altering every aspect \cite{-3,-2}. Computer-Assisted Pronunciation Training is one practical example, which may afford learners the opportunity to selectively target and practice specific accent features in a controlled and personalized manner. Conversely, applications such as movie dubbing require careful preservation of original phoneme durations to maintain audio-visual synchronization, thus inherently limiting the extent of accent modifications.

AC also addresses several practical communication challenges associated with accent differences. While widely recognized standard accents, such as Standard American English and Parisian French, tend to present minimal intelligibility issues within their respective regions or media contexts, cross-regional interactions can lead to significant comprehension difficulties \cite{73}, as well as prejudice and social stereotyping \cite{74}. These challenges are further amplified in speech technologies such as TTS and ASR, which often favor accents with abundant training data (for example, Standard American or British English). This imbalance results in performance disparities and recognition inaccuracies for underrepresented accents, causing these accents to be treated as out-of-distribution by such systems \cite{21}.

Applying AC within these technologies can improve intelligibility and reliability across diverse user groups \cite{2,9}. Additionally, personalized accented TTS systems developed through AC not only enhance intelligibility but also contribute to cultural relevance and authenticity in virtual interactions \cite{intercultural_awareness}. Furthermore, generating TTS voices with regional or minority accents through AC enables the electronic preservation of linguistic heritage beyond mere audio recordings \cite{south_voices}. This generative capability allows authentic accent reproduction for any given text, thereby playing a critical role in sustaining dialect diversity and supporting educational initiatives that emphasize cultural understanding.

\section{Challenges in Accent Conversion}
\label{sec:ac-challenges}

Despite significant progress in accent conversion research, the task remains inherently difficult due to the complex interplay between linguistic, acoustic, and identity-related factors. These difficulties arise both from the lack of suitable data and from the challenge of cleanly separating accent features from other aspects of speech. As a result, accent conversion faces two primary challenges: \textbf{Disentanglement Difficulty}: There are currently no publicly available parallel datasets where the same speaker produces speech in multiple accents \cite{23}. Consequently, AC algorithms often rely on weakly parallel datasets, in which different speakers utter the same content but in varying accents. This necessitates effective disentanglement of accent features from other speech characteristics, such as voice quality or linguistic content \cite{19, 16}. Given the inherent overlap between accent and content, where accent can be viewed as a form of content variation, it remains challenging to precisely isolate accent-specific features without inadvertently modifying other attributes. \textbf{Data Scarcity and Imbalance}: Publicly available weakly parallel datasets typically suffer from limited accent diversity, making it difficult for models to generalize well to previously unseen accents \cite{25}. In contrast, voice conversion (VC) datasets include thousands of unique speakers. Even when datasets have sufficient diversity, the volume of speech data for certain low-resource accents is often inadequate for models to effectively learn these accents, thus perpetuating performance gaps and limited representation.

\section{Related Tasks in Speech Processing}
\label{sec:related-tasks}

Accent Conversion is closely related to several speech processing tasks: \textbf{Voice Conversion (VC)}: Modifies speech from a source to sound like a target speaker by altering speaker-dependent attributes such as timbre and accent \cite{vc1, vc2}. This category also encompasses cross-lingual voice conversion (XVC), where speech is converted across languages while preserving speaker identity \cite{xvc1, xvc2}. AC can be viewed as a specialized instance of VC that focuses specifically on modifying accent characteristics. \textbf{Prosody and Style Transfer}: Modulates prosodic realization without changing lexical content or core speaker identity. This includes (i) speaking style such as clear or conversational speech \cite{style_trans1}; (ii) emotion such as happy, sad, or angry \cite{emo_trans1}; and (iii) intonation that adjusts pitch contour for sentence modality and focus patterns \cite{intonation_control}. \textbf{Accent Modeling}: Focuses on synthesizing accented speech directly, often under limited data conditions \cite{accent_modeling1, accent_modeling3}. To facilitate modeling, some approaches decompose accent into global and local components, enabling more structured control and improved generalization \cite{accent_modeling2}.

Each task contributes uniquely to advancing speech processing technologies, offering complementary insights and methodologies beneficial to AC research.

\section{Evaluation}
\label{sec:evaluation}

Accent conversion systems are typically evaluated along a common set of perceptual dimensions, which can be assessed using either objective or subjective methodologies. Objective metrics rely on automated signal- or model-based measurements, while subjective metrics depend on human listeners to directly assess perceptual attributes. Both approaches aim to characterize the same underlying dimensions but differ in how these attributes are quantified.

Across the literature, several core perceptual dimensions are commonly considered when evaluating AC systems, including overall audio quality, content preservation, similarity to the target accent, and speaker identity preservation. Depending on the experimental setup, these dimensions may be assessed using scalable objective metrics, perceptually grounded subjective evaluations, or a combination of both.

\subsection{Objective Evaluation of Accent Conversion}

Objective evaluation assesses the above dimensions using automated measurements derived from the signal itself or from pretrained models: \textbf{Audio Quality}: If reference audio is available, Mel Cepstral Distortion (MCD) can quantify differences in spectral envelope. Without reference, Signal-to-Noise Ratio (SNR) measures audio clarity. For distributional comparisons, metrics such as Fr\'{e}chet Audio Distance (FAD) \cite{FAD} can approximate perceptual quality. In addition, learned models that aim to predict subjective perceptual scores, such as Mean Opinion Score (MOS), have been proposed \cite{mosnet}. \textbf{Content Preservation}: ASR systems transcribe converted speech to text, enabling quantitative evaluation of linguistic consistency using word, phoneme, or character error rate (WER, PER, CER), as well as semantic similarity. These metrics vary in granularity, assessing content preservation from lexical to subphonemic levels. \textbf{Target Accent Similarity}: Historically, linguistic phoneme-based metrics such as ACCDIST \cite{47} characterized accent differences by comparing pronunciation patterns. More recently, accent classifiers are used to estimate similarity via confidence scores or embedding distance to a reference accent, although these metrics depend on the classifier's training data and inductive biases. \textbf{Speaker Identity Preservation}: Typically measured using embeddings learned from the speaker verification task, comparing extracted embeddings via cosine similarity or Euclidean distance.

Additional metrics such as pitch contour similarity, speaking rate consistency, or phoneme alignment can complement these evaluations, depending on whether the AC algorithm modifies these speech dimensions.

\subsection{Subjective Evaluation of Accent Conversion}

Subjective evaluation directly captures human perceptual judgments of the same core dimensions, offering an alternative and often more perceptually grounded assessment than automated metrics. Such evaluations are essential for assessing perceptual attributes that are difficult to reliably quantify using automated measures alone, and the choice of evaluation paradigm depends on the experimental setup, desired sensitivity, availability of reference audio, and cognitive load considerations.

These dimensions are commonly assessed using the following subjective rating paradigms: \textbf{Mean Opinion Score (MOS)}: Participants independently rate single audio samples, typically on a 1--5 scale. MOS is simple, scalable, and can be used to measure overall quality, speaker similarity, or similarity to the target accent depending on the evaluation prompt \cite{ITU-T-P800}. \textbf{Multiple Stimuli with Hidden Reference and Anchor (MUSHRA)}: Listeners rate multiple stimuli presented together on a continuous 0--100 scale, relative to a hidden reference. MUSHRA offers fine-grained discrimination and is suitable for sensitive evaluations of audio quality or accent similarity, but requires all samples to share identical linguistic content \cite{ITU-R-BS1534-1}. \textbf{A/B Testing}: Listeners compare two audio samples side by side and select the preferred one or the sample that sounds more similar to a given reference, such as the target accent, the original speaker, or a reference utterance. Judgments are typically binary (A or B) or based on a graded preference scale. This method has low cognitive load but provides less resolution than MUSHRA and MOS \cite{ITU-T-P800}.

In practice, AC studies often combine objective and subjective evaluations to obtain a comprehensive assessment of system performance. Objective metrics enable large-scale, automated comparisons and provide rapid preliminary insights, while subjective evaluations ensure perceptual validity by reflecting how listeners experience audio quality, speaker similarity, and accent authenticity.

\section{Datasets}
\label{sec:datasets}

Datasets specifically designed for AC tasks tend to be smaller compared to general-purpose speech datasets intended for TTS synthesis or ASR. Consequently, AC pipelines often rely on larger, unlabeled datasets for pretraining models or generating synthetic accented data \cite{23, 30}. For smaller datasets, it is particularly important to ensure phonetic balance, and many such datasets are explicitly designed to provide representative coverage of phonemes. Larger datasets, such as those sourced from audiobooks, naturally achieve phonetic diversity due to their size and scope. Common sampling rates for AC tasks range from 16 kHz, sufficient for intelligibility and ASR performance, up to 48 kHz for production-quality synthesis. Segmenting speech data at sentence boundaries and providing contextual information, including punctuation, supports learning sentence-level prosodic characteristics crucial for effective AC systems. Table~\ref{tab:datasets_summary} summarizes the characteristics of prominent English speech datasets frequently employed in accent conversion research.

\begin{table}[t]
\centering
\caption{Summary of English speech datasets used in accent conversion research. ``\#Acc.'' denotes the number of distinct accent categories. SR denotes sampling rate in kHz. Year indicates the initial or primary publication associated with each dataset. Statistics reflect the most recent publicly available version and may differ from those reported in the cited reference. Superscript $^s$ marks synthetic (TTS-generated) data. Accent descriptions follow the naming conventions of the original sources; for unannotated datasets, coverage is inferred from speaker demographics.}
\label{tab:datasets_summary}

\small
\setlength{\tabcolsep}{2.8pt}
\renewcommand{\arraystretch}{1.03}

\begin{tabularx}{\textwidth}{l c c c c c X}
\toprule
\textbf{Dataset} & \textbf{Year} & \textbf{SR} & \textbf{Hours} & \textbf{Spk.} & \textbf{\#Acc.} & \textbf{Accent coverage} \\
\midrule
\multicolumn{7}{l}{\textit{Unannotated (no accent labels)}} \\[2pt]
LibriLight    & 2020        & 16    & $\sim$60,000 & 7,439         & --   & Predominantly North American \\
LibriSpeech   & 2015        & 16    & $\sim$1,000  & 2,484         & --   & Predominantly North American \\
LibriTTS      & 2019        & 24    & 585          & 2,456         & --   & Predominantly North American \\
LibriTTS-R    & 2023        & 24    & 585          & 2,456         & --   & Predominantly North American \\
LJSpeech      & 2017        & 22.05 & 24           & 1             & --   & General American \\
\midrule
\multicolumn{7}{l}{\textit{Accent-annotated}} \\[2pt]
VCTK          & 2019        & 48    & 44           & 110           & 12   & British, Scottish, Irish, American, Canadian, Indian, Australian, and 5 others \\
Common Voice  & 2020        & 48    & $\sim$1,087  & $\sim$39,577  & Var. & American, British, Australian, Canadian, Indian, Irish, Filipino, Malaysian, etc. \\
SAA           & 2015        & Var.  & $\sim$10     & $\sim$2,100   & Var. & $>$200 native-language backgrounds represented \\
CMU Arctic    & 2004        & 16    & $\sim$7      & 7             & 4    & American, Canadian, Scottish, Indian \\
L2-Arctic     & 2018        & 44.1  & 27.1         & 24            & 6    & Hindi, Korean, Mandarin, Spanish, Arabic, Vietnamese \\
AccentDB      & 2020        & 48    & 9 + 11$^s$   & 8 + 15$^s$    & 9    & Indian (Bangla, Malayalam, Odia, Telugu, Metropolitan), American$^s$, British$^s$, Australian$^s$, Welsh$^s$ \\
UME-ERJ       & 2002        & 16    & $\sim$15     & 202           & 2    & Japanese, American \\
\bottomrule
\end{tabularx}

\vspace{2pt}
\raggedright\small
\textit{Datasets:} LibriLight \cite{LibriLight}, LibriSpeech \cite{LibriSpeech}, LibriTTS \cite{LibriTTS}, LibriTTS-R \cite{LibriTTS-R}, LJSpeech \cite{LJSpeech}, VCTK \cite{VCTK}, Common Voice \cite{CommonVoice}, Speech Accent Archive \cite{SAA}, CMU Arctic \cite{CMU-Arctic}, L2-Arctic \cite{L2-Arctic}, AccentDB \cite{AccentDB}, and UME-ERJ \cite{UME-ERJ}.
\end{table}

\section{Taxonomy of Accent Conversion Methods}
\label{sec:taxonomy}

The historical evolution of accent conversion methods reflects recurring scientific and data constraints that have shaped methodological design choices over time. Building on the three bottlenecks introduced in Section~\ref{sec:intro}, we review each paradigm by emphasizing how it addresses (i) the lack of parallel data and the need for correspondence, (ii) the requirement for reference-free conversion through disentanglement, and (iii) the challenge of scaling to low-resource and diverse accents.

\subsection{Early DSP-Based Methods}

Early AC methods used Digital Signal Processing (DSP) techniques to modify specific accent-related features, including phoneme articulation, intonation, and duration. Linear Predictive Coding (LPC) was employed to decompose speech into spectral envelope and excitation; modifying the envelope changed phoneme articulation, while the excitation was assumed to carry speaker-specific traits like fundamental frequency \cite{2}. Additionally, LPC residual signals were used to extract pitch contours for intonation adjustments \cite{-2}. Pitch-Synchronous Overlap and Add (PSOLA) modified suprasegmental prosodic features such as intonation and speech rate \cite{-3}. Another approach involved vowel formant shifting, utilizing manually computed formant statistics from target accents to reshape vowel quality \cite{0}.

To address the limitation in speaker identity preservation, \citet{3} proposed concatenative synthesis, selectively replacing only mispronounced segments using acoustic features such as Mel-Frequency Cepstral Coefficients (MFCCs) and articulatory features obtained from Electromagnetic Articulography (EMA). Although this partially improved identity preservation, synthesis quality suffered due to limited articulatory data. Generally, these methods required relatively little accented data and minimal computation but relied heavily on manual feature engineering and parallel utterances during training or inference. They also often produced unnatural artifacts such as robotic speech and struggled with robust speaker identity preservation. Despite these limitations, early DSP-based approaches found practical applications, most notably in foreign-language pronunciation training. They produced synthesized "golden speaker" utterances: target-accented speech matched to the learner's timbral characteristics, designed to facilitate native-like imitation \cite{2,-3,-2,3}. Another prominent application was multilingual TTS where methods based on accent conversion were used to correct pronunciation mismatches when synthesizing text containing multiple languages \cite{-1,1,0}.

\subsection{Data-Driven Methods and Improved Alignment}

With advances in machine learning, AC methods transitioned towards data-driven approaches. Ideally, parallel datasets in which the same speaker produces the same utterance in different accents would enable direct accent mapping; however, their scarcity led researchers to use weakly parallel datasets, where different speakers utter the same text. Because even matched utterances differ in timing, pronunciation, and prosodic realization, explicit alignment is required to establish frame-level or segment-level correspondence between source and target speech. Early alignment strategies employed Dynamic Time Warping (DTW) on acoustic features like MFCCs \cite{71}. This approach was appropriate for VC tasks where the goal was to fully imitate a target speaker, as it effectively transferred the entire speaker identity, including voice quality and accent. In applications like foreign-language pronunciation training, VC can be used to create a "golden speaker" by copying a non-native speaker's identity onto a native-accented utterance. In this setup, VC effectively serves as reference-based accent conversion, since it requires a native reference utterance during inference. However, the alignment method in \cite{71} could unintentionally transfer the non-native speaker's accent along with their voice quality onto the native utterance due to incorrect phoneme mappings driven by accent and timbre differences. To address this, \citet{4} introduced Vocal Tract Length Normalization (VTLN) prior to DTW, aligning utterances based on linguistic rather than acoustic similarity and thereby reducing accent-related alignment errors.

Subsequently, alignment methods shifted towards linguistic content using phoneme posteriorgrams extracted from speaker-independent ASR models, providing more accurate phoneme-level matching \cite{5}. Eventually, sequence-to-sequence (seq2seq) neural architectures emerged, eliminating the need for explicit alignment algorithms such as DTW \cite{11}. By learning soft alignments implicitly through attention mechanisms, these models improved the quality of source-target mappings and reduced artifacts associated with rigid, frame-level alignment procedures.

\subsection{Disentanglement of Accent, Content, and Prosody}

The limitation of requiring a reference utterance during inference motivated the development of reference-free AC methods. Effective reference-free accent conversion necessitates disentanglement of accent features from speaker identity and linguistic content in order to enable mapping from a source accent to a target accent without altering other speaker-specific attributes. Broadly, disentanglement techniques fall into the following categories: \textbf{Bottleneck methods}: These constrain information flow through a restricted intermediate representation (e.g., low-dimensional, discrete, or quantized bottlenecks such as Vector-Quantized Variational Autoencoders), encouraging the separation of partially overlapping attributes such as content, prosody, and speaker-related voice characteristics. \textbf{Supervised methods}: These guide disentanglement by explicitly predicting specific speech attributes, for example, predicting phonemes to isolate content or pitch contours to capture prosody. The chosen supervision target depends on which aspect is intended to be disentangled. \textbf{Adversarial methods}: These introduce auxiliary classifiers that penalize the presence of specific information in the learned representations (e.g., speaker-related traits, accent patterns, or prosodic cues such as pitch), thereby encouraging more factorized representations.

In practice, many AC systems combine these strategies. For instance, some methods use ASR models to extract speaker- and accent-independent linguistic features. Such ASR-based models employ phoneme prediction (supervised learning) together with bottleneck architectures or adversarial accent classifiers to enforce disentanglement, yielding accent-invariant content representations \cite{7}. Meanwhile, other approaches apply VQ-VAEs to derive prosody representations by combining dimensionality reduction with supervised pitch contour prediction \cite{10}.

\subsection{Efficient Accent Mapping with Rich-Resource Data}

After obtaining disentangled representations, effective mapping between source and target accents typically relies on weakly parallel data. However, because such data are scarce, particularly for low-resource accents, researchers have explored methods that leverage abundant native or non-accented speech to guide accent conversion models. \citet{7} use pretrained native TTS systems to provide latent-space guidance during training. These systems extract linguistic embeddings from native speech that steer the accent conversion model's latent space toward native-like pronunciation. While this approach helps achieve clearer articulation, it typically regenerates prosody entirely from the native reference, meaning that features such as emotion and intonation are not preserved. To introduce stronger supervision, \citet{30} proposed waveform-level alignment using synthetic native references. In this method, a native TTS system generates reference utterances from non-native inputs, creating an artificial parallel dataset. The converted utterances are then trained to match these synthetic references at the waveform level. This setup allows fine-grained control over accent modification, focusing primarily on phoneme articulation while maintaining the original prosodic characteristics. As a result, the converted speech preserves the speaker's identity even if it sounds less natively accented. Other works have employed cross-lingual transfer using multilingual TTS systems. Instead of collecting accented English data, the approach outlined in \citet{26} synthesized non-native English utterances by feeding transliterated English text into multilingual TTS models trained on other languages. The resulting speech naturally reflected the phonetic tendencies of the chosen language, enabling the construction of large-scale synthetic training pairs with native English utterances. More recently, researchers have explored pretraining strategies that use native speech directly, eliminating the need for intermediate native TTS generation. For example, \citet{23} pretrained an accent conversion module on native speech using masked token prediction over native semantic units, followed by fine-tuning on as little as 15 minutes of weakly parallel non-native data. This approach significantly reduced data requirements and achieved native-like output, although prosody was again regenerated by the pretrained native model, leading to the loss of emotional and intonational cues from the source speaker. Overall, these studies demonstrate a shift toward maximizing the utility of high-resource native datasets to improve accent mapping efficiency, each balancing the trade-off between accent modification strength and the preservation of speaker-specific traits.

\subsection{Many-to-Many and Any-to-Any Accent Conversion}

With advances in disentanglement techniques and improved strategies for handling low-resource accents, AC systems have evolved beyond simple one-to-one accent mappings toward more flexible configurations. These include many-to-one or any-to-one accent setups, where multiple non-native accents are converted into a single target accent (typical in FAC tasks) \cite{28,30,21,22,16}, as well as many-to-many accent frameworks that enable bidirectional or multi-accent conversion across diverse accent groups \cite{12,17,24}.

Many-to-many systems often represent target accents using discrete IDs \cite{24,17,12}, which allows the model to switch between known accents but limits generalization to unseen target accents. For more flexible generalization, such as many-to-any or any-to-any conversion, continuous accent embeddings are required. These embeddings are intended to encode actual accent characteristics rather than arbitrary labels.

Several studies have explored this direction, proposing to derive accent embeddings analogously to speaker embeddings in VC \cite{11,13,25}. They trained an accent classifier and extracted the hidden representations before the final classification layer to serve as accent embeddings. However, this approach, based solely on supervision, struggled to generalize to unseen accents for both source and target. The key limitations were the scarcity of training data for specific accents and insufficient diversity across accents, making it difficult for models to learn robust phonetic and prosodic features tied to each accent. To address these issues, \citet{29} proposed a more sophisticated disentanglement method using a Multi-Level VAE with vector quantization (VQ), which combined both supervised classification and bottleneck constraints to isolate accent-specific features more effectively. Despite these efforts, creating fully disentangled accent embeddings that generalize well remains an open problem, and true any-to-any accent conversion has yet to be fully achieved.

\section{Future Directions}
\label{sec:future}

Accent conversion has shown substantial progress, yet several critical challenges and opportunities remain open for future research: \textbf{Controllability and Speaker Identity Preservation}: Since accent constitutes a significant part of speaker identity, modifying accent inherently alters perceptual identity. Different accent aspects (e.g., phoneme articulation, intonation, speaking rate) affect identity preservation in varying degrees. Future AC systems should empower end-users with explicit control over identity preservation, allowing selective modification of specific accent attributes (e.g., modifying phoneme articulation without altering intonation) as well as continuous adjustment of the modification strength. Additionally, current methods typically regenerate prosodic features entirely from native speech patterns, thus disregarding important utterance-specific characteristics like emotional tone or expressive intention. Exploring prosody-aware conversion methods that preserve these nuanced elements represents a valuable direction. While recent works have begun to explore controllable accent manipulation and disentangled representations \cite{facfacodec, controllableAC2}, these approaches remain limited in scope, and more comprehensive frameworks for fine-grained and interpretable control are still needed. \textbf{Any-to-Any Accent Conversion}: As discussed earlier, capturing the full spectrum of accent variability through continuous embeddings remains challenging, primarily due to the lack of large and diverse datasets. Unlike timbre, which is relatively stable and can be inferred from only a few seconds of speech, accent depends on systematic patterns across diverse phonemic and prosodic contexts. Consequently, learning robust accent embeddings requires utterances that exhibit sufficient phonemic coverage and prosodic variation to represent the characteristic realizations of a given accent. Designing such datasets also draws on insights from linguistic research, particularly in determining minimal yet phonetically balanced text capable of capturing accent-defining features. Future research should therefore focus on developing robust disentanglement techniques and richer accent embeddings analogous to speaker embeddings. \textbf{Utilizing Unlabeled and Non-parallel Data}: Most contemporary accent conversion approaches rely heavily on weakly parallel datasets or synthetic data. However, weakly parallel datasets remain limited in diversity, while synthetic datasets often lack the complexity of real-world speech variations. Thus, future efforts should focus on effectively leveraging unlabeled and non-parallel speech datasets, enhancing techniques based on self-supervised representation learning, and better exploiting multilingual and cross-lingual speech resources to enable richer accent modeling and improve generalization to realistic scenarios. \textbf{Expanding Beyond English}: Modern accent conversion research predominantly concentrates on English accents, despite numerous other languages exhibiting significant regional accent and dialect diversity. Expanding accent conversion research beyond English is vital to promote linguistic inclusivity, facilitate global communication technologies, and support cultural preservation of regional dialects and accents. Future studies should explore cross-lingual methodologies, develop datasets and benchmarks for additional languages, and adapt current AC methods to address the unique linguistic characteristics encountered in languages other than English. \textbf{Integrating Lexical Variations}: Although this survey defines accent narrowly as pronunciation, capturing segmental and suprasegmental speech features, related linguistic aspects such as lexical choices, syntactic structures, and even region-specific spelling conventions can also signal a speaker's regional or cultural background. Current AC methods primarily focus on conversion at the speech realization level, modifying pronunciation within the audio signal. Extending AC systems to incorporate language modeling that captures and optionally modifies these higher-level features presents a promising direction. This could enable conversion not only of how something is said, but also what is said and how it is written, enhancing the authenticity and sociolinguistic richness of the converted speech.

\section{Conclusion}
\label{sec:conclusion}

Accent conversion has seen remarkable development over recent decades, driven by increasing industrial and academic interest in improving global and cross-cultural communication. This survey presented a systematic overview of the evolution of accent conversion methodologies, tracing progress from early DSP-based feature manipulations to contemporary neural architectures capable of sophisticated accent transformations. By outlining key advances across different methodological stages, ranging from data-driven alignment techniques to disentanglement-based and many-to-many systems, it provides a coherent view of how the field has matured, highlighting emerging challenges and future directions in achieving flexible and high-fidelity accent conversion.

Widely used speech datasets were reviewed and common subjective and objective evaluation methods were outlined, emphasizing their complementary roles in robust algorithm assessment. Finally, by identifying critical open problems and future research avenues, the survey aimed to stimulate further innovation. Addressing these research gaps will be pivotal in realizing practical, inclusive, and scalable accent conversion systems that meet the diverse communication needs of our increasingly interconnected world.

\printcredits

\section*{Declaration of competing interest}
The authors declare that they have no known competing financial interests or personal relationships that could have appeared to influence the work reported in this paper.

\section*{Data availability}
No data was used for the research described in the article.

\section*{Funding}
This research did not receive any specific grant from funding agencies in the public, commercial, or not-for-profit sectors.

\bibliographystyle{unsrtnat}
\bibliography{tacl2018}

\end{document}